\def\la{\mathrel{\hbox{\rlap{\hbox{\lower4pt\hbox{$\sim$}}}\hbox{$<$}}}}
\def\ga{\mathrel{\hbox{\rlap{\hbox{\lower4pt\hbox{$\sim$}}}\hbox{$>$}}}}

\documentstyle[preprint,aps]{revtex}
\tightenlines
\begin{document}

\title {COSMOLOGICAL MAGNETIC FIELDS}
\author{Angela V. Olinto
 \footnote{e-mail: olinto@oddjob.uchicago.edu}}

\address{Department of Astronomy \& Astrophysics, and \\
Enrico Fermi Institute, The University of Chicago,\\
5640 S. Ellis Av., Chicago, IL 60637}
\maketitle
\begin{abstract}
We discuss the evolution of cosmological
magnetic fields from the early universe to 
the present.  We review different scenarios for magnetogenesis 
in the early universe and follow the subsequent  evolution of these
fields as the universe recombines. We then focus on the role primordial
fields play after recombination in the seeding of
stellar and galactic fields and the formation of structure.  Cosmological
magnetic fields in the intergalactic medium trace the turbulent history
of the universe and may contain fossils 
 of the early universe. We conclude by discussing   
observational probes of  cosmological magnetic fields including the study
of extragalactic cosmic rays.

(Invited talk at the 3rd RESCEU International Symposium on ``Particle
Cosmology'', University of Tokyo, Nov. 10-13, 1997.)
\end{abstract}

\section{Introduction}

Present cluster and galactic magnetic fields are observed to be in
energy equipartition with the gas and the cosmic rays in these systems.
Although the origin of these fields is still unclear, they are most likely
the result of strong amplification of seed fields present in protogalactic
clouds. These seed fields may be primordial in nature, being leftover from
the early universe, or they may have been created when the first structures
formed after recombination. 

Determining if the origin of cluster and galactic fields are primordial or
post-recombination is a difficult task, since the amplification by many
orders of magnitude in these virialized systems overwhemls any signature of
an early magnetic phase. In contrast, the presence of fields away from
clusters and galaxies can more easily be related to the early history of
cosmological magnetic fields. However, magnetic fields in
extragalactic and extra-cluster regions are quite difficult to observe. At
present, synchrotron maps and Faraday rotation measurements show some
evidence for magnetic fields on inter-cluster scales, while Faraday
rotation of high redshift objects give only upper limits to fields on
cosmological scales \cite{1}.  

Improved direct observations of extragalactic fields,
such as high resolution Faraday rotation maps, together with alternative
methods that have been recently proposed, such as the study of
extragalactic cosmic rays,  may help determine the origin and evolution of
cosmological magnetic fields.  If large-scale fields are present
throughout the universe today, their structure and
spectrum should have clearer signatures of their origin. 

Understanding the history
of magnetic fields  will help understand the dynamics  of the
early universe and the formation
of the first stars and galaxies. In particular, magnetic fields may be 
as relevant to galaxies as they are to the Sun. In addition,  
large-scale magnetic fields also trace  the hydrodynamic evolution of the
universe, as magnetic fields are fossil records of shocks, outflows, and
winds, as well as of early universe processes. In this respect, it is
important to be able to differentiate between the present structure of
extragalactic fields formed by primordial fields and  those formed by 
pollution of galactic outflows.

In the following sections, we follow the history of magnetic fields
from magnetogenesis in the early universe to the present  and
discuss possible ways of observing the unknown structure of
cosmological magnetic fields. We start by reviewing  early universe
magnetogenesis. We then discuss the evolution and damping of magnetic
fields up to recombination. Fields present at recombination    can play a
role on the formation of galaxies and stars by preserving fluctuations
below the Silk mass  and by generating density perturbations after
recombination.  Finally, magnetic fields on scales of clusters and larger
play an important role on the origin and propagation of
extremely-high--energy cosmic rays. As future experiments try to  address
the origin and nature of EHECRs, the structure of magnetic fields in the
intergalactic medium around the Local Group will be simultaneously
addressed.

\section{Magnetic Fields in the Early Universe}

Historically, the study of magnetic field generation in the early
universe was motivated by the search for the origin of galactic
large scale fields. For instance, typical spiral galaxies have large scale
fields of about  a few $\mu$G ($\equiv 10^{-6}$ G), coherent over
the plane of their disks. (The Milky way has $B_{MW}
\simeq 3 \mu$G.) These galactic fields could have originated from a
relatively large primordial seed field amplified by the collapse of
the galaxy or by a much smaller seed field that is greatly
amplified by a galactic dynamo. These two possibilities give very
different estimates for the field needed to seed the galactic
field. (In the absence of sources of currents, magnetohydrodynamic
equations are linear in 
$B$, thus,  seed fields are necessary for the growth of magnetic fields
with time in a plasma.)

If the galactic dynamo is efficient at amplification, the seed
field can be as low as
$B_{seed}\simeq 10^{-23}$ G. This value is estimated assuming that
the galactic dynamo amplified the field present at the newly formed
galaxy by a factor of $\sim e^{30}$ corresponding to about 30
dynamical timescales or complete revolutions since the galaxy
formed. After the dynamo amplification factor is taken into
account, the present galactic field  requires that the newborn
galaxy had a field of
$\sim 10^{-19}$ G.  Finally, we can  estimate  the
primordial field needed to explain present galactic fields by
assuming that the newborn galaxy amplified a frozen-in cosmological
field  as the protogalactic primordial gas collapsed to present
galactic densities. As the gas collapses, the magnetic field increases
as  $B \propto
\rho^{2/3}$. This gives  primordial fields of about
$B_{seed} \simeq 10^{-23}$ G for a density increase of $10^6$
between the galactic density and the average density in the
universe.  If a galactic dynamo is not efficient at amplification
the primordial field requirements increase to $B_{seed}
\simeq 10^{-9} - 10^{-12}$ G \cite{2}.

At present, a successful galactic dynamo model remains elusive. Dynamo
models work  well for the Earth  where fluctuations about the large
scale field are small compared to the average field. This hierarchy
of scales does not hold in stars and galaxies, making the dynamo
problem more challenging. A compass would not be very useful on
stars or galaxies since the fluctuation fields are of the same order of
or stronger than  the large scale coherent field. Here
we follow the two extremes of primordial seeding,
with and without dynamo amplification.

The challenge for early universe magnetogenesis depends on the
magnitude of the needed seed fields. 
 A useful way of comparing the magnitude of fields at
different times is to write the magnetic energy density,
$\rho_B \equiv B^2 /8
\pi$,  in units of the radiation background energy density, 
$\rho_{\gamma}\equiv \pi^2 g_* T^4/30 $ where $g_*$ is the number of
degrees of freedom and $T$ is the temperature of the cosmic background
radiation (CBR). The present CBR temperature implies 
$\rho_{\gamma} \simeq (4 \mu{\rm G})^2/ 8 \pi$. Both energy
densities redshift the same way with the expansion of the universe 
and can be compared as comoving quantities ($B
\propto a^{-2}$ and
$\rho_B \propto a^{-4}$, where $a(t)$ is the cosmic scale factor).
In these units the field strength required to seed  galactic fields
with an efficient galactic dynamo translates into
$\rho_B \simeq 10^{-34} \rho_{\gamma}$, while the primordial seed
without dynamo amplification requires $\rho_B \simeq 10^{-14}
\rho_{\gamma}$.

Once a primordial field is generated, it  redshifts with the
expansion of the universe and damps due to viscous processes. The most
significant damping occurs at decoupling due to photon viscosity (see
\S2.3). Magnetic diffusion is usually insignificant due to the high
conductivity of the plasma. The residual ionization of the
universe today is sufficient to guarantee that field diffusing is
insignificant on scales above 
$\sim $ A.U.\cite{3,4} Here we discuss fields on scales larger
than a parsec, with emphasis on galaxy scales which correspond to comoving
scales of about 1 Mpc.

\subsection{Primordial Magnetogenesis}

A number of authors have proposed scenarios for generating  seed
fields in the early universe \cite{4}--\cite{13}. These models 
make use of the few out-of-equilibrium epochs of the
pre-recombination era. Phase-transitions are usually necessary,
with models ranging from inflation (when the universe had
temperatures, $T  \ga \, 10^{16}$ GeV) to the QCD transition
($T  \sim 100$ MeV). 

Standard inflationary models give rise to insignificantly small
vector perturbations in contrast to the observable scalar and tensor
perturbations. The simplest inflationary models give $\rho_B \sim
10^{-104} \rho_{\gamma}$ which is  too small to act as a seed field.
Reasonable seed fields can be generated if one breaks
electromagnetic gauge invariance,\cite{3} or changes the
gravitational couplings \cite{7}. String cosmology may also give
rise to primordial fields \cite{8}  but the magnitude of the
effect is still under debate \cite{9}. 

Although somewhat contrived, inflation based models for seed field
generation have the advantage that large scale fields can be easily
generated due to the exponential expansion during vacuum
domination. On the other hand, observations of the cosmic
background radiation (CBR) seem to strongly constrain these
models, leaving little room for their viability \cite{10}. 

Several scenarios have been proposed  below the GUT scale such as
during the Electroweak and the QCD transitions \cite{11,12,13}. These
are usually based on out-of-equilibrium processes during first or
second order phase-transitions where shocks, interfaces, and
turbulent motions may generate significant fields through battery
and amplification mechanisms.  

Models based on phase-transitions other than inflation are limited
by causality and result in  less power on large scales. In order to compare
different models,  we use the definitions in \cite{14} where the Fourier
transform, ${\bf \tilde B}({\bf k})$,  of the magnetic field at a
point in space, ${\bf B}({\bf x})$, can be described by a scalar
power spectrum ${\tilde B}^2(k)$ when  ${\bf B}({\bf x})$ is 
assumed to be a random, homogeneous, and isotropic  field. The
total magnetic  energy density   is then given by:    
$$ \rho_B \equiv {1\over 2}\int {\tilde B}^2(k) d^3k \ . $$ 

The spectrum generated by different models can then be described by
power laws on length scales above the horizon, $H_{pt}^{-1}$, (or
wavenumbers below
$ 2 \pi H_{pt}$) of the particular phase transition. During
radiation domination, the horizon for any given phase-transition is
much smaller than the typical intergalaxy separation,  ($H_{pt}^{-1}
\ll $ Mpc), thus a power law behavior is a good description of the
spectrum on the scales of interest. (The last transition of
interest is the QCD transition when the universe had a horizon that
corresponds to a comoving scale of  $H_{QCD}^{-1} \simeq $ 1 pc
today. Earlier transitions have much smaller horizon scales.)
Thus, we can write  
${\tilde B}^2(k) = A k^n $, where the constant $A$ is related to
the total magnetic field energy density, $\rho_B$, the spectral index,
$n$,   and the cutoff scale of the  spectrum,  $k_{\rm
max}$, as  $A \simeq (n+3) 8 \pi \rho_B/ k_{\rm max}^{n+3}$.

Causality limits the spectrum of fields generated at transitions
during radiation dominated epochs to have spectral index $n \ge 0$,
while inflationary models can, in principle, generate fields with
negative spectral indices, i.e., more power is present on large scales.

Independent of the specific mechanism for magnetic field
generation, we can estimate the ability of phase transitions to
seed galactic fields, given the limit on the total energy available
for magnetic fields, the power index limit due to causality, and
the maximum cutoff scale given by the horizon.  The maximum energy
density in magnetic fields that any scenario can generate is
$\rho_B = \rho_{\gamma}$. This is an overestimate since
hydrodynamical processes usually lead to magnetic energies in
equipartition with plasma motions and $\rho_B \simeq \rho_{\rm
plasma} {\rm v}^2 \la \, 0.1 \,
\rho_{\gamma}$ for most phase transitions. The cutoff together with  the
maximum energy  limit the amplitude of the
spectrum for a given spectral index. For the electroweak transition
the cutoff is $k_{EW} \simeq 2 \pi H_{EW} \simeq 2 \pi/ {\rm 10 AU}$
while for the QCD transition the cutoff is 
$k_{QCD} \simeq 2 \pi H_{QCD} \simeq 2 \pi/ {\rm pc}$.

The best case scenario for  large scales fields generated in phase
transitions is the white noise spectrum with
$n=0$ \cite{5}. In this case, we can estimate the maximum
strength of the magnetic field on Mpc scales by choosing a window
function that extracts the contribution of the field spectrum on
these scales. Following  \cite{14}, for $n=0$ the average field on a
scale $l$ is:
$${\bar B}(l) \le \pi^{3/4} \, \sqrt{6 \rho_B \over (l  k_{\rm
max})^{3}}$$ This model-independent estimate gives upper limits
for fields generated in the  electroweak transition  of ${\bar
B}_{EW}({\rm Mpc}) \  \la \ 10^{-22}$ G and for the QCD transition
${\bar B}_{QCD} ({\rm Mpc}) \ 
\la $ $  10^{-16}$ G. Therefore, neither transition can  seed the
galactic field without  dynamo amplification since the generated
fields are much smaller than the needed $B_{\rm seed}
\sim 10^{-9} - 10^{-12}$ G.  

\subsection{Evolution up to Recombination}

After primordial fields are generated, they redshift with the
expansion  of the universe, being frozen into the plasma for most of the
early universe's  history. Although magnetic diffusion is  
insignificant,  during certain epochs in the early universe,  
magnetic field energy is converted into heat through the damping of
magneto-hydrodynamic (MHD) modes \cite{16}.  This damping is caused
by dissipation in the fluid due to the finite mean free path of
photons and neutrinos. The result of these fluid viscosities is the
efficient damping of MHD modes similar to the Silk damping of
adiabatic density perturbations.

The evolution of MHD modes, such as  fast and slow magnetosonic,
and Alfv\'en waves, in the presence of viscous and heat conducting
processes can be studied in both the radiation diffusion and the
free-streaming regimes of the early universe. Fluid viscosities
damp cosmic magnetic fields from prior to the epoch of neutrino
decoupling up to recombination. Similar to the case of sound waves
propagating in a demagnetized plasma, fast magnetosonic waves are
damped by radiation diffusion on all scales smaller than the
radiation diffusion length. The characteristic damping scales are
the horizon scale at neutrino decoupling ($M_{\nu} \approx 10^{-4}
M_{\odot}$ in baryons) and the Silk mass at recombination
($M_{\gamma} \approx 10^{13} M_{\odot}$ in baryons).  In contrast,
the oscillations of slow magnetosonic and Alfv\'en waves get
overdamped in the radiation diffusion regime, resulting in
frozen-in magnetic field perturbations. Further damping of these
perturbations is possible only if before recombination the wave
enters a regime in which radiation free-streams on the scale of the
perturbation. The maximum damping scale of slow magnetosonic and
Alfv\'en modes is always smaller than or equal to the damping scale
of fast magnetosonic waves, and depends on the magnetic field
strength and its direction relative to the wave vector \cite{16}.

The dissipation of magnetic energy into heat  during neutrino
decoupling weakens big bang nucleosynthesis constraints on the
strength of magnetic fields present during nucleosynthesis. The
observed element abundances require that $\rho_b \la \rho_{\gamma}
/ 3$ during  nucleosynthesis \cite{17,18}. Even if processes prior to
neutrino decoupling generate magnetic fields with  $\rho_b \simeq
\rho_{\gamma}$ initially, neutrino damping causes the magnetic energy
to decrease substantially relative to that of radiation by the time of
nucleosynthesis. This ensures that most magnetic field configurations
generated prior to neutrino decoupling satisfy big bang
nucleosynthesis constraints. 

Dissipation during recombination weakens the ability of
primordial magnetic fields to generate  galactic magnetic
fields or density perturbations. Since a sizable fraction of the
energy in magnetic field fluctuations is erased up to the Silk scale,
it becomes even more difficult to produce the observed galactic fields
without dynamo amplification. Models which generate primordial
fields in sub-horizon scales during phase transitions are
particularly constrained  since these models have more  power on
small scales where damping is most efficient.

Although Alfv\'en and slow magnetosonic modes also undergo
significant damping, these modes survive on scales    
smaller than the damping scale for fast magnetosonic modes and
magnetic energy can be stored   on scales
well below the Silk mass. The survival of these modes help the
seeding of galactic fields by preserving magnetic energy on scales of
interest to galaxy formation, and may also be of significance to the
formation of structure on relatively small scales. In particular,
these modes may be responsible for fragmentation of early structures
and the seeding of early star or galaxy formation.

\subsection{Early Dynamos}

The nature of hydrodynamic flows as the universe recombines is an
important factor in the evolution of cosmological magnetic fields. If
there are velocity flows present as the universe evolves, these flows
may have enough helicity  to set up early dynamos. Magnetic energy
can be regained at the expense of the kinetic energy of the flow and 
primordial fields may be amplified to the necessary level to seed
galactic fields \cite{12,13}. Some simulations  of pre-recombination MHD
systems show  dynamo behavior \cite{15,19}.  However, viscous damping due
to photon and neutrino decoupling damp velocity flows as well as MHD
modes through recombination.  If velocity fields exist at
recombination, the most likely scale for driving the flow will be
close to the Silk mass. Since the couplings in this problem are
non-linear, it is possible that the driving at large scales cascades
down to small scales and back to large scales, amplifying fields 
during and  after recombination. Constrains on such pre-recombination
flows should be studied in the light of recent  CBR observations. 

Alternatively, the collapse of density perturbations generated by
inflation and imprinted in a non-baryonic cold dark matter
component may generate the necessary seed field and amplification. 
 Numerical simulations of the growth of density perturbations after
recombination show the  formation of shocks that can generate seed
fields , which can then be amplified by large scale dynamos \cite{20}.

\section{Magnetic Fields in the late Universe}

\subsection{Galaxy Formation} 

While magnetic fields are recognized as central agents in regulating
the dynamics of star formation and the general interstellar medium
in galaxies, it is generally assumed that they do not play a
significant role during the epoch of galaxy formation. However, if
fields of the order of
$10^{-9}$G to $10^{-12}$G are present in protogalactic clouds, this
view should be questioned. Depending on the spectrum of such
cosmological fields, structure formation can be initiated by
magnetic fields \cite{14,23}. 

Magnetic fields may act as seeds for density
perturbations. The evolution of density perturbations and  
peculiar velocities seeded by primordial magnetic fields give rise
to a steep spectrum of density perturbations\cite{14}, namely $P(k)
\sim k^4$. This spectrum is too steep to account for the observed
large-scale structure of the universe, thus,   magnetic fields
alone cannot reproduce the observed clustering on large scales. 
 On the other hand,  magnetic fields do generate  small-scale
structure shortly after recombination, even if the rms magnetic
field on intergalactic scales is as small as $10^{-12}$ Gauss
today. Thus, magnetic fields may provide a natural source of
(scale-dependent) bias of the luminous baryonic matter with respect
to the dark matter in the universe. 

 Another consequence of primordial magnetic fields is to add power
on small scales to the primordial density perturbation spectrum, a
welcome ingredient for models of structure formation which lack
small-scale power, such as tilted cold dark matter, mixed dark
matter, and hot dark matter.

\subsection{Pollution of the IGM}

When trying to measure the magnitude of cosmological magnetic
fields today, the most significant observations are those of
magnetic fields on the largest scales and away from virialized
systems such as galaxies and clusters of galaxies. Thus,
observations of magnetic fields in the intergalactic medium are the
most helpful in understanding the origin of cosmological magnetic
fields.  

Intergalactic and intercluster magnetic fields of significant
magnitudes have been observed,\cite{1,24} but their
interpretation is not unique. One of the questions is the ability of
galaxies to pollute the intergalactic medium and if
``pollution'' fields can  be differentiated from 
 primordial fields. Galactic outflows may pollute a significant
fraction of the IGM \cite{25}  depending on  models of
galaxy outflows, galaxy formation, and cosmological evolution. 

With the discovery of the highest energy cosmic rays  and the
possibility of studying the sources of these events in the future,
the question of whether an extragalactic magnetic field exists
outside of clusters has gained a new observational tool. 

\section{Observing Cosmological Magnetic Fields}

\subsection{Direct Observations}

Of great relevance to understanding the history of cosmological
magnetic fields  are observations of intergalactic magnetic fields
and magnetic fields at high redshifts. Reports of Faraday rotation
associated with high-redshift Lyman-$\alpha$ absorption systems
(see, e.g., \cite{1}) suggest that dynamically significant magnetic
fields (of order $\mu$G) may be present in condensations at high
redshift. Together with observations of strong magnetic fields in
clusters,\cite{24}  these observations support the idea that
magnetic fields play a dynamical role in the evolution of structure
and maybe present throughout the universe.

Present Faraday rotation measurements of intergalactic fields using
the emission from high-redshift quasars place limits of
$\la 10^{-9}$ G for fields with Mpc reversal scales, and $\la
10^{-11}$G for fields coherent on the present horizon scale. 
Future   measurements may help determine the strength of galactic
magnetic fields at high redshifts as well as the presence of
significant fields in the IGM today.

\subsection{CBR Signatures of B}

Future observations of CBR anisotropies   by MAP and
PLANCK will be able to detect the acoustic
Doppler peaks from sound waves at recombination and polarization. If
magnetic fields of significant magnitude are present at
recombination, they polarize the CBR photons\cite{21} and generate
signatures of the different MHD modes \cite{22}. The precise
signatures  are likely to depend on the nature of the mode, since
each MHD mode evolves differently through recombination \cite{16}.

\subsection{Extragalactic Cosmic Rays}

The  detection of extremely high-energy cosmic rays (EHECRs) has
triggered considerable interest in the  origin and nature of these
particles.  To date, more than 60 cosmic ray events with energies
above $\sim 5 \times 10^{19}$ eV  have been observed by experiments such
as Haverah Park, Fly's Eye, and  AGASA. The Fly's Eye experiment has
recorded a $3 \times 10^{20}$ eV event, the highest energy event so
far. These EHECRs  most likely  originate from extragalactic
sources and their spectrum and spatial as well as temporal
distributions are  affected by the presence of extragalactic
magnetic fields. 

The study of EHECRs is closely related to the study of cosmological
magnetic fields \cite{26}.  Charged particles of energies  $ \sim
10^{20}\,$eV can be deflected significantly in cosmic magnetic
fields. Thus,  detailed information on the structure of
 extragalactic magnetic fields may be contained in the time,
energy, and arrival direction distributions of charged
extremely-high energy particles emitted from powerful discrete
sources. In this context, the clustering among EHECRs suggested by
recent AGASA data is very encouraging and its confirmation would
have important consequences for understanding the nature and origin
of cosmological  magnetic fields as well as for EHECRs \cite{27}.

Another signature of cosmological magnetic fields is in the shape
of the spectrum of  $\gamma$-rays secondaries to EHECRs. The
$\gamma$-ray spectrum depends on  synchrotron losses versus  
inverse-Compton regeneration of the electromagnetic cascade. This
signature  is most sensitive to extragalactic magnetic fields
around the Faraday rotation limit of $\sim 10^{-9}$G \cite{28}. Weaker
fields may be detected TeV electromagnetic cascades \cite{29,30}.

\section{Conclusion}

We followed some of the history of cosmological magnetic fields. At
each step of this history, new questions arise. Magnetogenesis in
phase transitions alone cannot generate galactic fields, but is
there a pre-recombination dynamo or does the amplification occur as  
galaxies form? Decoupling damps fast magnetosonic waves, but do
Alfv\'en and slow magnetosonic modes play a role on scales below
the Silk mass? Protogalactic shocks may generate small seed fields,
but do these get amplified on large scales?

These are some of the questions that future observations of large
scale magnetic fields will address. In addition to 
traditional direct observations, the study of extremely high-energy
cosmic rays will play an important role in probing cosmological fields
in a 50 Mpc volume around us.

\section{Acknowledgments}  We acknowledge the support of the US
Department of Energy and the National Science Foundation at the
University of Chicago.


\begin{references}

\bibitem{1}       P. P. Kronberg, {\it Rep. Prog. Phys.} {\bf 57}, 325
(1994).
\bibitem{2}      R.M. Kulsrud,   in {\it Galactic and Intergalactic
Magnetic Fields}, ed.\ R.\ Beck, P.P.\ Kronberg, \& R.\ Wielebinski
(Dordrecht: Kluwer), p.\ 527 (1990).
\bibitem{3} 
M. S. Turner and L. M. Widrow,
{\it Phys. Rev. D} {\bf 30} {2743} {1988}.
\bibitem{4}      B. Cheng and A. V. Olinto,
{\it Phys. Rev. D} {\bf 50} {2421} {1994}.
\bibitem{5}      C. J. Hogan, {\it Phys. Rev. Lett.}{\bf 51}{1488}{1983}.  
\bibitem{6} 
E. R. Harrison, {\it MNRAS} {\bf 147}, 279
(1970); E. R. Harrison, {\it MNRAS} {\bf 165}, 185 (1973);  W. D.
Garretson, G. B. Field, and S. M. Carroll,
{\it Phys. Rev. D}{\bf 46}, 5346 (1992);  A. D. Dolgov,
{\it Phys. Rev. D}{\bf 48}, 2499 (1993);  A. D. Dolgov and J. Silk,
{\it Phys. Rev. D}{\bf 47}, 3144 (1993). 
\bibitem{7} 
B. Ratra, {\it Ap. J} {\bf 391}, L1 (1992);  B.
Ratra, {\it Phys. Rev. D} {\bf 45}, 1913 (1992); 
\bibitem{8} 
M. Gasperini, M. Giovannini, and G. Veneziano,
{\it Phys. Rev. Lett.} {\bf 75},  3796 (1995).
\bibitem{9}  
D. Lemoine and M. Lemoine, {\it Phys. Rev. D}
{\bf 52}, 1955 (1995);
\bibitem{10}
D. Lemoine, PhD Thesis (1995)
\bibitem{11} 
T. Vaschaspati, {\it Phys. Lett. B} {\bf 265}, 258
(1991);  R. H. Brandenberger, A.-C. Davis, A. M. Matheson, and M.
Trodden, {\it Phys. Lett.  B} {\bf 293}, 287 (1992);  A. P. Martin
and A.-C. Davis, {\it Phys. Lett. B} {\bf 360}, 71 (1995);  T. W.
B. Kibble and A. Vilenkin, {\it Phys. Rev. D } {\bf 52}, 679 (1995);
 J. Quashnock, A. Loeb, and D. N. Spergel, {\it Ap. J.} {\bf 344},
L49 (1989).
\bibitem{12} 
G. Baym, D. B\"odecker, and L. McLerran, {\it Phys.
Rev. D} {\bf 53}, 662 (1996);
\bibitem{13}
G. Sigl, K. Jedamzik, and A. V. Olinto, {\it Phys.
Rev. D} {\bf 55}, 4582 (1997).
\bibitem{14}  
E. Kim,  A. V. Olinto, \& R. Rosner,  {\it Ap. J.},
467 (1996).
\bibitem{15} 
K. Enqvist, P. Olensen, {\it Phys. Lett. B} {\bf 329},
195 (1994).
\bibitem{16}  K. Jedamzik, V. Katalinic, and A. V. Olinto, {\it
Phys. Rev. D}, in press (1998).
\bibitem{17} 
P. Kernan, G. Starkman, and T. Vachaspati, {\it Phys. Rev. D} 
{\bf 54}, 7202 (1996).
\bibitem{18} 
B. Cheng, A. V. Olinto, D. Schramm, and J. Truran,
{\it Phys. Rev. D} {\bf 54}, 4714 (1996).
\bibitem{19}
A. Brandenburg, K. Enqvist, and P. Olesen, {\it Phys.
Lett.} {\bf B391}, 395 (1997).
\bibitem{20} 
R.  Kulsrud, D. Ryu, R. Cen,  and J. P.  Ostriker,
{\it Ap. J.} (1997).
\bibitem{21} 
A. Loeb and A.   Kosowsky, {\it  Ap..J.} {\bf 469} 1
(1996).
\bibitem{22}  
J. Adams, U. Danielsson, D. Grasso, and H. Rubinstein,
{\it Phys.Lett.} {\bf B388} 253 (1996).
\bibitem{23}
I. Wasserman,  {\it  Ap. J.}, {\bf 224}, 337 (1978).
\bibitem{24}
K.-T. Kim,   P.C. Tribble, and P.P. Kronberg, {\it
Ap. J.}, {\bf 379}, 80 (1991).
\bibitem{25}
P.  Kronberg,   and H.  Lesch,    in "The Physics of
Galactic Halos" eds. H. Lesch, R-J Dettmar, U. Mebold and R.
Schlickeiser, Berlin: Akademie Verlag (1996).
\bibitem{26} 
M. Lemoine, A. V. Olinto, G. Sigl, and D. Schramm,  
{\it Ap. J. Lett.} (1997)
\bibitem{27} 
G. Sigl, A. V. Olinto, and M. Lemoine, {\it Phys.
Rev. D} 
 (1997).
\bibitem{28} 
S.  Lee, A. V. Olinto, and G. Sigl,  {\it Ap. J.
Lett.} {\bf 455}, L1 (1995).
\bibitem{29} R. Plaga, {\it Nature} {\bf 374}, 430 (1995).
\bibitem{30} 
E. Waxman and P. Coppi, {\it Ap. J.} {\bf 464}, L75 (1996).

\end{references}
\end{document}